\documentclass[
  prb, %
  aps,
  eqsecnum,
  notitlepage,
  nofootinbib,
  superscriptaddress,
  10pt
]{revtex4-2}

\usepackage[utf8]{inputenc}
\usepackage[T1]{fontenc}
\usepackage{lmodern}

\usepackage{graphicx}
\usepackage{amsmath,amssymb,amsfonts,bm}
\usepackage{xcolor}

\usepackage{orcidlink}

\usepackage{hyperref}
\hypersetup{
  colorlinks=true,
  linkcolor=blue,
  citecolor=blue,
  urlcolor=blue,
  pdfstartview=FitV,
  breaklinks=true
}

\begin{document}

\title{Time-Reversal Symmetry in Artificial Topological Graphene Rings}


\author{J. A. da Silva Neto\,\orcidlink{0009-0007-1945-362}}\email{amarocz@gmail.com}
\affiliation{Escola Cidadã Integral e Técnica Nicéa Claudino Pinheiro,
58900-000, Cajazeiras, PB, Brazil.}


\author{G. Q. Garcia\,\orcidlink{0000-0003-3562-0317}}
\email{gqgarcia99@gmail.com}
\affiliation{Centro de Ci\^encias, Tecnologia e Sa\'ude,
Universidade Estadual da Para\'iba,
58233-000, Araruna, PB, Brazil.}


\author{A. M. de M. Carvalho\,\orcidlink{0009-0006-3540-0364}}
\email{alexandre@fis.ufal.br}
\affiliation{Instituto de F\'{\i}sica,
Universidade Federal de Alagoas,
57072-970, Macei\'o, AL, Brazil.}


\author{C. Furtado\,\orcidlink{0000-0002-3455-4285}}
\email{furtado@fisica.ufpb.br}
\affiliation{Departamento de F\'isica,
Universidade Federal da Para\'iba,
58051-970, Jo\~ao Pessoa, PB, Brazil.}


\begin{abstract}
We study an artificial graphene ring described by the low-energy
two-valley Dirac Hamiltonian with a uniform mass term. When the ring
is threaded by an Aharonov--Bohm flux, the problem reduces to a
second-order radial equation solved by Bessel functions. We give the
exact secular equation and, for narrow rings, closed expressions for
the spectrum and the persistent current. We then add a uniform
magnetic field perpendicular to the plane. The radial problem is
solved with confluent hypergeometric functions, and the persistent
current and magnetization are written as derivatives of a secular
determinant from which the spurious roots have been removed. Besides
the usual orbital coupling, the field enters through a
valley-dependent term that comes from the Dirac structure and lifts
the valley degeneracy. A weak-field expansion, accurate up to
corrections of order $B^4$, gives this valley splitting explicitly
and separates the diamagnetic and paramagnetic contributions to the
susceptibility. Finally, we work out what time reversal implies for
the spectrum. Physical time reversal, which exchanges the valleys,
leads to an exact relation between levels at opposite fields. The
effective single-valley time reversal, on the other hand, is broken
by the mass term and by the boundary conditions, yet leaves no trace
in the zero-field spectrum. Numerical results for the spectrum, the
persistent current, and the magnetization illustrate the analysis.
\end{abstract}

\maketitle

\section{Introduction}
Topology shapes the electronic properties of many condensed-matter
systems. Dirac materials~\cite{banerjee}, among them graphene and
related two-dimensional crystals such as hexagonal boron nitride
(h-BN) and silicene~\cite{Novoselov,Geim,castro,liu,gian}, have become
a versatile platform for exploring this connection~\cite{wehling}.
Dirac-like states are also at the heart of topological
insulators~\cite{hasan,Qi,Chen}, and they can be engineered in
artificial electronic systems~\cite{Gomes,Polini}. What these systems
share is that their low-energy excitations obey an effective Dirac
equation. In graphene, the nearly linear dispersion around the Dirac
points makes the carriers behave as massless Dirac fermions, which
has turned graphene into a natural testing ground for confined
geometries such as quantum dots~\cite{Silvestrov,Ponomarenko,Stampfer},
quantum rings~\cite{Xavier,Amaro1,Amaro2}, and
antidots~\cite{Shen}.

Quantum rings are particularly well suited to studying interference.
A magnetic flux threading the ring imprints an Aharonov--Bohm
phase~\cite{Aharonov} on the electronic states even where the
magnetic field itself vanishes. The resulting flux dependence of the
spectrum produces persistent currents, one of the clearest signatures
of phase coherence in mesoscopic rings~\cite{Byers,Cheung,BakkePersistent}.

Symmetry adds another layer. Experiments on graphene quantum dots
revealed confinement-induced spectral features that prompted the
study of effective time-reversal-symmetry breaking in
graphene~\cite{Ponomarenko,Wurm}. In chaotic or diffusive systems,
whether time-reversal symmetry (TRS) is present or absent largely
determines the transport properties~\cite{Wurm}. An external magnetic
field breaks TRS outright, but graphene calls for more care, because
physical time reversal maps one valley onto the other.

The neutrino billiard of Berry and Mondragon~\cite{Berry} offers a
useful analogy here. Its Dirac Hamiltonian is closely related to the
low-energy Hamiltonian of a single graphene valley, for which an
effective intravalley TRS can be broken without any magnetic field.
In graphene, the symmetry class that results depends on intervalley
scattering and on the nature of the edges and of the
disorder~\cite{McCann,Morpurgo,Fujita,Wimmer}. Suitable boundary
conditions can even produce an effective mass confinement analogous
to the infinite-mass condition of the Berry--Mondragon
model~\cite{Berry}.

In this paper we study an artificial graphene ring described by a
low-energy Dirac Hamiltonian. We begin with a single-layer ring
threaded by an Aharonov--Bohm flux and obtain its spectrum and
persistent current. We then add a uniform magnetic field
perpendicular to the plane and follow how it changes the spectrum
and the magnetic response. Our main interest is the way the field
lifts the valley degeneracy and how this shows up in the persistent
current and the magnetization. Along the way, we separate the
relations between energy levels that follow from physical time
reversal from those that would follow from the effective
single-valley time reversal.

The paper is organized as follows. Section~\ref{sec:hamiltonian}
introduces the effective Hamiltonian and derives a second-order
equation for the spinor components. In Sec.~\ref{sec:AB-effect} we
solve the Aharonov--Bohm ring and compute the spectrum and the
persistent current. Section~\ref{sec:uniform-magnetic-field} adds the
uniform field and presents the exact secular equation, the
weak-field expansion, and the magnetic response. The time-reversal
properties of the spectrum are discussed in Sec.~\ref{sec:TRS}, and
Sec.~\ref{sec:conclusions} summarizes our conclusions.

\section{Model Hamiltonian}
\label{sec:hamiltonian}

We describe the low-energy states of single-layer graphene in the
continuum Dirac approximation. Close to the two inequivalent valleys,
$K$ and $K'$, the quasiparticles carry a sublattice and a valley
degree of freedom, on which the Pauli matrices $\sigma_i$ and
$\tau_i$ act, respectively. With an external vector potential
$\mathbf{A}$ and a spatially uniform mass term $V$, the effective
Hamiltonian reads~\cite{Wurm}
\begin{equation}
H
=
v(p_x-eA_x)\sigma_x\otimes\tau_z
+
v(p_y-eA_y)\sigma_y\otimes\tau_0
+
V\sigma_z\otimes\tau_0,
\label{eq:effective-Hamiltonian}
\end{equation}
where $v$ is the Dirac velocity, $V$ is the mass that opens a gap in the spectrum, and $\tau_0$ is the $2\times2$ identity in valley space. The vector potential $\mathbf{A}=(A_x,A_y)$ describes the magnetic configurations considered below. The annular geometry and the boundary conditions that confine the carriers radially are introduced in Sec.~\ref{sec:AB-effect}. It is convenient to rewrite Eq.~(\ref{eq:effective-Hamiltonian}) with Dirac matrices obeying the Clifford algebra $\{\gamma^\mu,\gamma^\nu\} = 2g^{\mu\nu}I_4$, where the metric tensor is given by $g^{\mu\nu} = \mathrm{diag}(1,-1,-1,-1)$, with $\mu,\nu=0,1,2,3$. A representation adapted to the sublattice and valley structure is
\begin{equation}
\gamma^0 = \sigma_z\otimes\tau_0,\ \
\gamma^1 = i\sigma_y\otimes\tau_x,\ \
\gamma^2 = i\sigma_y\otimes\tau_y, \ \text{and}\ 
\gamma^3 = i\sigma_y\otimes\tau_z.
\end{equation}
Defining $\beta=\gamma^0$, these matrices satisfy the following relations:
\begin{equation}
\sigma_x\otimes\tau_z = \beta\gamma^3,\ \
\sigma_y\otimes\tau_0 = \gamma^1\gamma^2\gamma^3,\ \ \text{and}\ \
\sigma_z\otimes\tau_0 = \beta.
\end{equation}
And therefore, the Hamiltonian in Eq.~(\ref{eq:effective-Hamiltonian}) can be equivalently written as
\begin{equation}
H
=
v(p_x-eA_x)\beta\gamma^3
+
v(p_y-eA_y)\gamma^1\gamma^2\gamma^3
+
V\beta.
\label{eq:effective-hamiltonian}
\end{equation}

In order to solve the equation~\eqref{eq:effective-hamiltonian}, we must consider the eigenvalues equation for stationary states, $H\Psi=E\Psi$. Thus, considering the natural units $\hbar=c=1$, we can write a {\it ansatz} for the solution
\begin{equation}
\Psi(t,x,y)
=
e^{-iEt}
\begin{pmatrix}
\eta(x,y)\\
\chi(x,y)
\end{pmatrix},
\label{eq:spinor-ansatz}
\end{equation}
where $\eta(x,y)$ and $\chi(x,y)$ are the two sublattice components, each of them a spinor in valley space:
\begin{equation}
\eta
=
\begin{pmatrix}
\eta_K\\
\eta_{K'}
\end{pmatrix},
\qquad
\chi
=
\begin{pmatrix}
\chi_K\\
\chi_{K'}
\end{pmatrix}.
\end{equation}
The four-component spinor $\Psi$ thus carries both degrees of
freedom, and $\tau_z$ distinguishes the valleys,
\begin{equation}
\tau_z\eta
=
\begin{pmatrix}
\eta_K\\
-\eta_{K'}
\end{pmatrix},
\qquad
\tau_z\chi
=
\begin{pmatrix}
\chi_K\\
-\chi_{K'}
\end{pmatrix}.
\end{equation}
Now, let us introduce the kinetic momentum operators, so that
\begin{equation}
\Pi_i
=
p_i-eA_i,
\qquad
i=x,y,
\label{eq:kinetic-momentum}
\end{equation}
with $p_j=-i\hbar\partial_j$, and then the eigenvalue problem becomes the pair of coupled equations
\begin{subequations}
\begin{eqnarray}
(E - V)\eta &=& v(\Pi_x\tau_z - i\Pi_y)\chi; \label{eq:coupled-eta}\\
(E + V)\chi &=& v(\Pi_x\tau_z + i\Pi_y)\eta, \label{eq:coupled-chi}
\end{eqnarray}
\end{subequations}
or in the coordinate representation, we have that
\begin{subequations}
\begin{eqnarray}
(E - V)\eta &=& - iv\tau_z\frac{\partial\chi}{\partial x} - evA_x\tau_z\chi - v\frac{\partial\chi}{\partial y} + ievA_y\chi; \label{eq:coordinate-eta}\\
(E+V)\chi &=& -iv\tau_z\frac{\partial\eta}{\partial x} - evA_x\tau_z\eta + v\frac{\partial\eta}{\partial y} - ievA_y\eta. \label{eq:coordinate-chi}
\end{eqnarray}
\end{subequations}

Assuming that $V$ is spatially uniform, we eliminate $\chi$ by combining Eqs.~\eqref{eq:coupled-eta} and (\ref{eq:coupled-chi}). From Eq.~\eqref{eq:coupled-chi}, we have that $\chi$ to be eliminated for $E\neq-V$. Thus, substituting this expression into Eq.~(\ref{eq:coupled-eta}) and using the spatial uniformity of $V$, we obtain
gives
\begin{align}
(E^2-V^2)\eta
={}&
v^2
(\Pi_x\tau_z-i\Pi_y)
(\Pi_x\tau_z+i\Pi_y)\eta
\nonumber\\
={}&
v^2
\left(
\Pi_x^2+\Pi_y^2-e\tau_z B_z
\right)\eta,
\label{eq:second-order-operator}
\end{align}
where
\begin{equation}
B_z
=
\frac{\partial A_y}{\partial x}
-
\frac{\partial A_x}{\partial y},
\label{eq:magnetic-field}
\end{equation}
and we have used
\begin{equation}
[\Pi_x,\Pi_y]
=
ieB_z.
\label{eq:momentum-commutator}
\end{equation}
Expanding the kinetic momenta, we arrive at the second-order differential equation
\begin{align}
&v^2\frac{\partial^2\eta}{\partial x^2}
+
v^2\frac{\partial^2\eta}{\partial y^2}
-
iev^2
\left(
\frac{\partial A_x}{\partial x}
+
\frac{\partial A_y}{\partial y}
\right)\eta
\nonumber\\
&\quad
-
2iev^2
\left(
A_x\frac{\partial\eta}{\partial x}
+
A_y\frac{\partial\eta}{\partial y}
\right)
-
e^2v^2(A_x^2+A_y^2)\eta
\nonumber\\
&\quad
+
ev^2\tau_z B_z\eta
+
(E^2-V^2)\eta
=
0,
\label{eq:second-order-expanded}
\end{align}
Writing the vector potential as $\mathbf{A}=(A_x,A_y)$, Eq.~(\ref{eq:second-order-expanded}) can be expressed in the compact form
\begin{align}
v^2\nabla^2\eta
&-
iev^2
(\boldsymbol{\nabla}\cdot\mathbf{A})\eta
-
2iev^2
\mathbf{A}\cdot\boldsymbol{\nabla}\eta
-
e^2v^2|\mathbf{A}|^2\eta
\nonumber\\
&+
ev^2\tau_z B_z\eta
+
(E^2-V^2)\eta
=
0.
\label{eq:second-order-compact}
\end{align}
Note that the equation~\eqref{eq:second-order-compact} governs the low-energy dynamics for both magnetic configurations studied here. Its next-to-last term couples the magnetic field to the valley degree of freedom through $\tau_z$. For a pure Aharonov--Bohm flux this term vanishes wherever the carriers can go, since $B_z=0$ there; it becomes important once a uniform field is applied to the ring. We treat the two cases in turn.

\section{Aharonov--Bohm ring}
\label{sec:AB-effect}

We first consider the Aharonov--Bohm effect~\cite{Aharonov} for the Hamiltonian~(\ref{eq:effective-hamiltonian}), with a magnetic flux along the $z$ axis. Working in the Coulomb gauge,
\begin{equation}
\boldsymbol{\nabla}\cdot\mathbf{A}=0,
\label{eq:coulomb-gauge}
\end{equation}
we rewrite Eq.~(\ref{eq:second-order-compact}) in polar coordinates $(\rho,\varphi)$. The flux is confined to the hole of the ring, $\rho<a$, which the carriers cannot reach. In the annulus, the vector potential
is~\cite{Aharonov}
\begin{equation}
\mathbf{A}
=
\frac{\phi}{2\pi\rho}
\hat{\boldsymbol{\varphi}},
\label{eq:AB-vector-potential}
\end{equation}
where $\phi$ is the flux. The vector potential does not vanish in
the annulus, but the field does,
\begin{equation}
B_z=0,
\qquad \rho\geq a,
\label{eq:AB-field-outside}
\end{equation}
so the valley-dependent term of Eq.~(\ref{eq:second-order-compact}) plays no role in this configuration. From the equation~\eqref{eq:AB-vector-potential},
\begin{equation}
    \mathbf{A}\cdot\boldsymbol{\nabla} = \frac{\phi}{2\pi\rho^2} \frac{\partial}{\partial\varphi}\ \ \text{and}\ \ |\mathbf{A}|^2 = \frac{\phi^2}{4\pi^2\rho^2}.
\end{equation}
and, in natural units, the flux quantum is
\begin{equation}
\phi_0
=
\frac{2\pi}{e}.
\label{eq:flux-quantum}
\end{equation}

Substituting these expressions into Eq.~(\ref{eq:second-order-compact}), we obtain the following expression
\begin{align}
&
v^2\frac{\partial^2\eta}{\partial\rho^2}
+
\frac{v^2}{\rho}\frac{\partial\eta}{\partial\rho}
+
\frac{v^2}{\rho^2}
\frac{\partial^2\eta}{\partial\varphi^2}
\nonumber\\
&\quad
-
\frac{2iv^2}{\rho^2}
\frac{\phi}{\phi_0}
\frac{\partial\eta}{\partial\varphi}
-
\frac{v^2}{\rho^2}
\left(\frac{\phi}{\phi_0}\right)^2\eta
+
(E^2-V^2)\eta
=
0,
\label{eq:AB-wave-equation}
\end{align}
or, more compactly,
\begin{equation}
v^2
\left[
\frac{\partial^2}{\partial\rho^2}
+
\frac{1}{\rho}\frac{\partial}{\partial\rho}
+
\frac{1}{\rho^2}
\left(
\frac{\partial}{\partial\varphi}
-i\frac{\phi}{\phi_0}
\right)^2
\right]\eta
+
(E^2-V^2)\eta
=
0.
\label{eq:AB-wave-equation-compact}
\end{equation}
We look for solutions of the form
\begin{equation}
\eta(\rho,\varphi)
=
e^{il\varphi}
\begin{pmatrix}
R_{+}(\rho)\\
R_{-}(\rho)
\end{pmatrix},
\label{eq:angular-ansatz}
\end{equation}
where $R_{+}\equiv R_{K}$ and $R_{-}\equiv R_{K'}$ are the radial
functions of the two valleys and
\begin{equation}
l=0,\pm1,\pm2,\ldots
\label{eq:angular-quantum-number}
\end{equation}
is the orbital angular quantum number of $\eta$. Because $\eta$ is
written in the Cartesian sublattice basis, a single-valued wave
function requires $l$ to be an integer. The half-integer eigenvalues
$j=l\pm1/2$ of the total angular momentum
$J_z=L_z+\frac{1}{2}\sigma_z\otimes\tau_z$ belong to the full
four-component spinor, whose other sublattice component carries the
orbital number $l\pm1$. The flux leaves this quantization untouched:
it only shifts the angular term in
Eq.~(\ref{eq:AB-wave-equation-compact}).

Inserting Eq.~(\ref{eq:angular-ansatz}) into
Eq.~(\ref{eq:AB-wave-equation-compact}) gives two decoupled radial
equations, one for each $R_s(\rho)$, $s=\pm$,
\begin{equation}
\frac{d^2 R_s(\rho)}{d\rho^2}
+
\frac{1}{\rho}
\frac{d R_s(\rho)}{d\rho}
-
\frac{\xi^2}{\rho^2}R_s(\rho)
+
\nu^2 R_s(\rho)
=
0.
\label{eq:AB-radial}
\end{equation}
The valleys decouple because the Hamiltonian has no intervalley
terms, and, since $B_z=0$, both obey the same radial equation.
The parameters in Eq.~(\ref{eq:AB-radial}) are
\begin{equation}
\xi
=
l-\frac{\phi}{\phi_0},
\label{eq:AB-xi}
\end{equation}
and
\begin{equation}
\nu^2
=
\frac{E^2-V^2}{v^2}.
\label{eq:AB-nu}
\end{equation}
With $\mu=\nu\rho$, the equation~\eqref{eq:AB-radial} becomes Bessel's
differential equation,
\begin{equation}
\frac{d^2R_s(\mu)}{d\mu^2}
+
\frac{1}{\mu}
\frac{dR_s(\mu)}{d\mu}
+
\left(
1-\frac{\xi^2}{\mu^2}
\right)R_s(\mu)
=
0,
\label{eq:AB-bessel-mu}
\end{equation}
whose general solution is
\begin{equation}
R_s(\rho)
=
A_{\xi}J_{|\xi|}(\nu\rho)
+
B_{\xi}Y_{|\xi|}(\nu\rho),
\label{eq:AB-bessel-solution}
\end{equation}
with $J_{|\xi|}$ and $Y_{|\xi|}$ the Bessel functions of the first
and second kind.

\begin{figure}[!htb]
    \centering
    \includegraphics[width=0.75\textwidth]{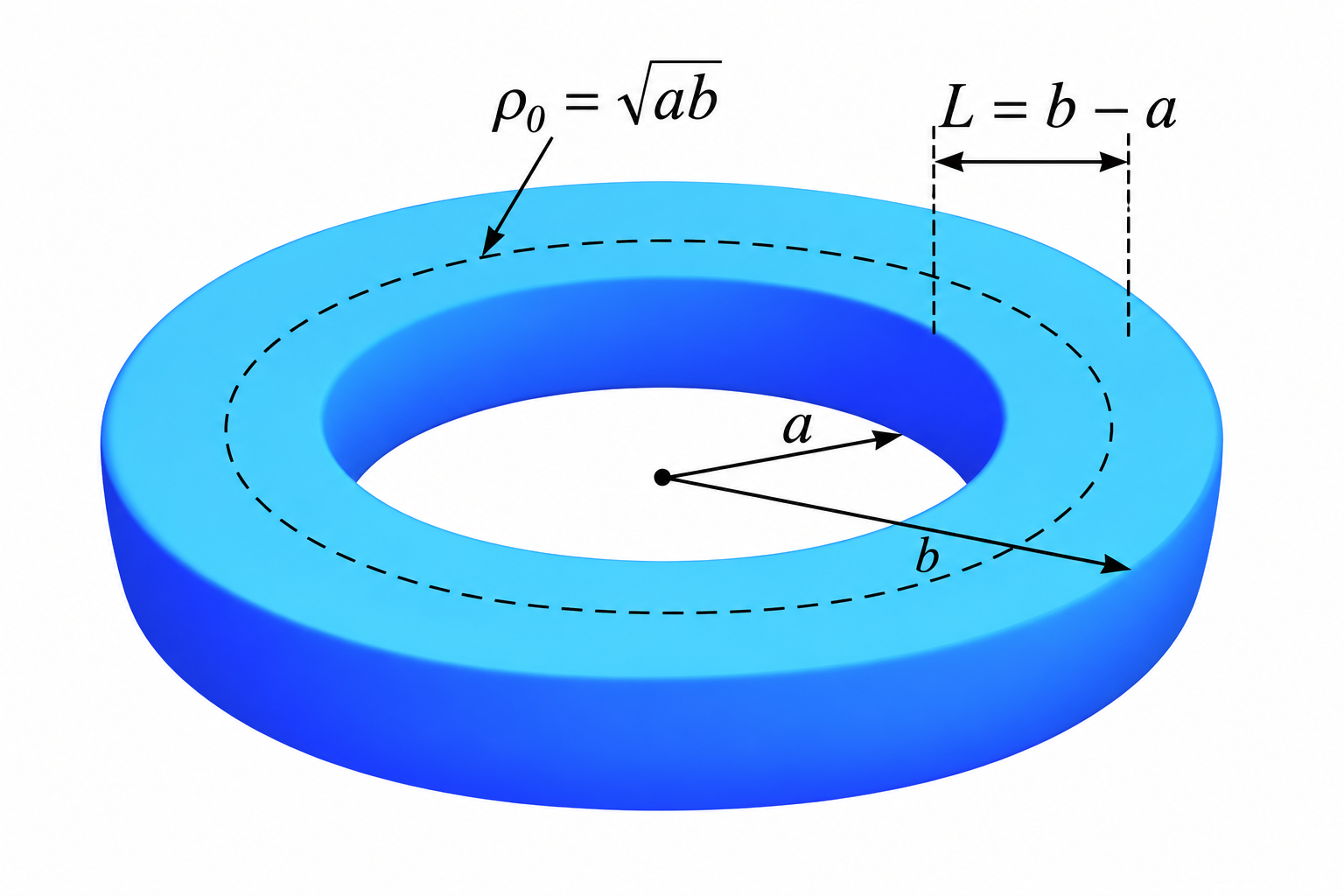}
    \caption{Sketch of the graphene ring, with inner radius $a$, outer
    radius $b$, width $L=b-a$, and geometric mean radius
    $\rho_0=\sqrt{ab}$.}
    \label{fig1}
\end{figure}

Following the artificial topological ring of Kumagai and
Ohno~\cite{kumagai}, we confine the particle to an annulus of inner
radius $a$ and outer radius $b>a$, with width and geometric mean
radius
\begin{align}
L &= b-a,
\label{eq:ring-width}\\
\rho_0 &= \sqrt{ab},
\label{eq:ring-mean-radius}
\end{align}
as sketched in Fig.~\ref{fig1}. We require the radial function to
vanish at both edges,
\begin{equation}
R_s(a)=R_s(b)=0,
\label{eq:ring-boundary-conditions}
\end{equation}
which fixes the allowed radial modes. Only the sublattice component
$\eta$ is forced to vanish, so Eq.~(\ref{eq:ring-boundary-conditions})
is a zigzag-type condition. The radial current density is bilinear
in $\eta$ and $\chi$ and therefore vanishes at both edges, as it
should. This condition differs from the infinite-mass condition of
the Berry--Mondragon model~\cite{Berry}, which ties the two
sublattice components together at the boundary. Note also that
eliminating $\chi$ through Eq.~(\ref{eq:coupled-chi}) leaves out the
sector $E=-V$, in which $\eta\equiv0$ and
$(\Pi_x\tau_z-i\Pi_y)\chi=0$. These states are the continuum
counterparts of zigzag edge states, and we do not consider them
here. In short, the mass $V$ sets the gap, while the boundary
conditions provide the radial confinement.

Applied to Eq.~(\ref{eq:AB-bessel-solution}), the boundary
conditions give
\begin{equation}
\begin{pmatrix}
J_{|\xi|}(\nu a) & Y_{|\xi|}(\nu a)\\
J_{|\xi|}(\nu b) & Y_{|\xi|}(\nu b)
\end{pmatrix}
\begin{pmatrix}
A_{\xi}\\
B_{\xi}
\end{pmatrix}
=
\begin{pmatrix}
0\\
0
\end{pmatrix}.
\label{eq:AB-boundary-matrix}
\end{equation}
Nontrivial solutions require a vanishing determinant, which yields
the secular equation
\begin{equation}
J_{|\xi|}(\nu a)Y_{|\xi|}(\nu b)
-
J_{|\xi|}(\nu b)Y_{|\xi|}(\nu a)
=
0.
\label{eq:AB-secular}
\end{equation}
Calling $\nu_{\xi,n}$ its $n$th root, the radial eigenfunction is
\begin{equation}
R_s(\rho)
=
\mathcal{N}
\left[
J_{|\xi|}(\nu_{\xi,n}\rho)
-
Y_{|\xi|}(\nu_{\xi,n}\rho)
\frac{
J_{|\xi|}(\nu_{\xi,n}b)
}{
Y_{|\xi|}(\nu_{\xi,n}b)
}
\right],
\label{eq:AB-radial-eigenfunction}
\end{equation}
with $\mathcal{N}$ a normalization constant. The roots $\nu_{\xi,n}$ give the energies through Eq.~(\ref{eq:AB-nu}). An approximate analytical spectrum follows in the regime
\begin{equation}
\nu a\gg 1,
\qquad
\nu b\gg 1,
\label{eq:AB-asymptotic-regime}
\end{equation}
where the large-argument expansions of the Bessel functions apply.
They require the order to stay small compared with the argument,
$|4\xi^2-1|\ll 8\nu a$, and we use them below for narrow rings,
$L\ll\rho_0$.

To first order in $(\nu a)^{-1}$,
\begin{align}
J_{|\xi|}(\nu a)
\simeq{}&
\sqrt{\frac{2}{\pi\nu a}}
\left[
\cos\left(
\nu a-\frac{|\xi|\pi}{2}-\frac{\pi}{4}
\right)
\right.
\nonumber\\
&\left.
\qquad
-
\frac{4\xi^2-1}{8\nu a}
\sin\left(
\nu a-\frac{|\xi|\pi}{2}-\frac{\pi}{4}
\right)
\right],
\label{eq:AB-J-asymptotic}
\end{align}
and
\begin{align}
Y_{|\xi|}(\nu a)
\simeq{}&
\sqrt{\frac{2}{\pi\nu a}}
\left[
\sin\left(
\nu a-\frac{|\xi|\pi}{2}-\frac{\pi}{4}
\right)
\right.
\nonumber\\
&\left.
\qquad
+
\frac{4\xi^2-1}{8\nu a}
\cos\left(
\nu a-\frac{|\xi|\pi}{2}-\frac{\pi}{4}
\right)
\right],
\label{eq:AB-Y-asymptotic}
\end{align}
and similarly at $\rho=b$. Substituting these expansions into
Eq.~(\ref{eq:AB-secular}) and using
Eqs.~(\ref{eq:ring-width}) and~(\ref{eq:ring-mean-radius}), we find
for $L\ll\rho_0$
\begin{equation}
\nu^2
\simeq
\left(\frac{n\pi}{L}\right)^2
+
\frac{4\xi^2-1}{4\rho_0^2},
\label{eq:AB-nu-narrow-ring}
\end{equation}
where $n=1,2,3,\ldots$ labels the radial modes. In this narrow-ring regime~\cite{kumagai}, the radial term $(n\pi/L)^2$ dominates the curvature correction for any fixed $|\xi|$, including $\xi\to0$. The correction is negative for $|\xi|<1/2$: at $\phi=0$, for example, the $l=0$ state lies $1/(4\rho_0^2)$ below the straight-channel value $(n\pi/L)^2$. This is a geometric zero-point effect of the annular confinement~\cite{kumagai}. Combining the equations~\eqref{eq:AB-nu-narrow-ring}, \eqref{eq:AB-nu} and~\eqref{eq:AB-xi}, the spectrum is
\begin{equation}
E_{n,l}
=
\pm
\left\{
V^2
+
v^2
\left[
\left(\frac{n\pi}{L}\right)^2
+
\frac{
4\left(
l-\frac{\phi}{\phi_0}
\right)^2-1
}{
4\rho_0^2
}
\right]
\right\}^{1/2},
\label{eq:AB-energy-spectrum}
\end{equation}
with the positive and negative branches describing electron-like
and hole-like states.

The persistent current follows from the Byers--Yang
relation~\cite{Byers,Cheung}. At zero temperature it is the flux
derivative of the total energy,
\begin{equation}
I
=
-\frac{\partial U}{\partial\phi}
=
-\sum_{(n,l)\in\mathrm{occ}}
\frac{\partial E_{n,l}}{\partial\phi},
\label{eq:AB-persistent-current-definition}
\end{equation}
where
\begin{equation}
U
=
\sum_{(n,l)\in\mathrm{occ}} E_{n,l}
\label{eq:AB-total-energy}
\end{equation}
is the energy of the occupied single-particle states for a fixed
number of carriers. For electron doping we sum over the occupied
positive-energy states; the filled negative-energy sea would require
its own reference prescription, which we do not discuss. Using the following relations:
\begin{equation}
\xi
=
l-\frac{\phi}{\phi_0},
\qquad
\frac{\partial\xi}{\partial\phi}
=
-\frac{1}{\phi_0},
\label{eq:AB-xi-flux-derivative}
\end{equation}
and $\phi_0=2\pi/e$, each occupied state contributes
\begin{equation}
I_{n,l}
=
\pm
\frac{e}{2\pi}
\frac{v^2\xi}{\rho_0^2}
\left\{
V^2
+
v^2
\left[
\left(\frac{n\pi}{L}\right)^2
+
\frac{4\xi^2-1}{4\rho_0^2}
\right]
\right\}^{-1/2},
\label{eq:AB-current-state}
\end{equation}
to the current, the upper (lower) sign referring to the positive
(negative) branch. Since the valleys are degenerate here and the
Hamiltonian~(\ref{eq:effective-Hamiltonian}) does not act on spin,
a fully occupied level $(n,l)$ carries a degeneracy factor
$g_sg_v=4$. The natural current scale is $I_0=ev/(2\pi\rho_0)$,
although each contribution also depends on the radial energy and on
$l$.

The total current is
\begin{equation}
I
=
\sum_{(n,l)\in\mathrm{occ}}
I_{n,l}.
\label{eq:AB-persistent-current}
\end{equation}
The flux enters only through $\xi=l-\phi/\phi_0$, which satisfies
\begin{equation}
\xi(l+1,\phi+\phi_0)
=
\xi(l,\phi).
\label{eq:AB-flux-periodicity}
\end{equation}
Hence
\begin{equation}
E_{n,l+1}(\phi+\phi_0)
=
E_{n,l}(\phi),
\label{eq:AB-spectrum-periodicity}
\end{equation}
and the spectrum as a whole is periodic in $\phi$ with period
$\phi_0$, up to a relabeling of $l$. Any occupation prescription
that respects this relabeling gives a persistent current with the
same period.

\subsection*{Numerical results}
\begin{figure}[!htb]
\centering
\includegraphics[width=0.5\textwidth]{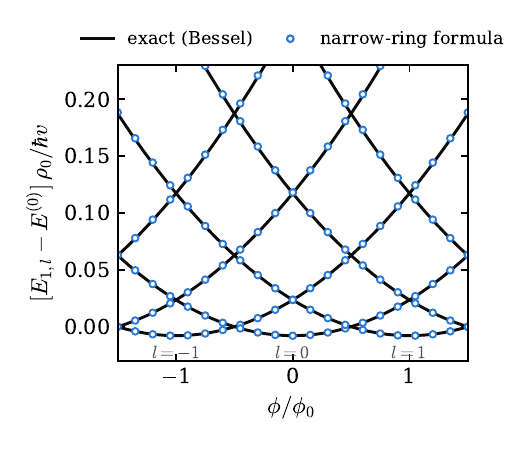}
\caption{Energy levels of the lowest radial mode ($n=1$) versus the
Aharonov--Bohm flux for $l=-3,\ldots,3$, with $L=0.2\rho_0$ and
$V=2\hbar v/\rho_0$. Solid lines: exact roots of
Eq.~(\ref{eq:AB-secular}); circles: narrow-ring formula,
Eq.~(\ref{eq:AB-energy-spectrum}). Energies are measured from
$E^{(0)}=[V^2+(\pi\hbar v/L)^2]^{1/2}$.}
\label{fig:AB-spectrum}
\end{figure}

For the numerical examples we measure lengths in units of $\rho_0$
and energies in units of $\hbar v/\rho_0$, and take $L=0.2\rho_0$
and $V=2\hbar v/\rho_0$. For $\rho_0=50$~nm and $v=10^6$~m/s, the
energy unit is $\hbar v/\rho_0\simeq13$~meV.
Figure~\ref{fig:AB-spectrum} compares the exact levels of the lowest
radial mode, $n=1$, obtained by solving
Eq.~(\ref{eq:AB-secular}) numerically, with the narrow-ring
formula~(\ref{eq:AB-energy-spectrum}). Energies are measured from
$E^{(0)}=[V^2+(\pi\hbar v/L)^2]^{1/2}$. Each angular channel traces a
nearly parabolic branch centered at $\phi=l\phi_0$, and neighboring
branches cross at half-integer $\phi/\phi_0$. Even though
$L/\rho_0=0.2$ is not especially small, the narrow-ring formula
stays within $4\times10^{-3}\,\hbar v/\rho_0$ of the exact result
over the whole range, that is, within $2\%$ of the flux-dependent
part of the spectrum.

\begin{figure}[!htb]
\centering
\includegraphics[width=0.5\textwidth]{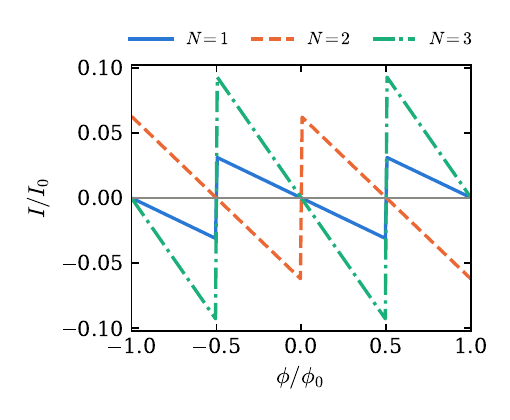}
\caption{Persistent current at $T=0$ versus the Aharonov--Bohm flux
for $N=1,2,3$ electrons in one spin and valley sector, in units of
$I_0=ev/(2\pi\rho_0)$. Parameters as in Fig.~\ref{fig:AB-spectrum}.}
\label{fig:AB-current}
\end{figure}

Figure~\ref{fig:AB-current} shows the zero-temperature persistent
current for $N$ electrons filling the lowest positive-energy levels
of one spin and valley sector; because the valleys are degenerate
here, the total current follows by including the degeneracy factor.
The current has the familiar sawtooth shape, with period $\phi_0$
and jumps at the level crossings of Fig.~\ref{fig:AB-spectrum}. For
odd $N$ the jumps sit at half-integer $\phi/\phi_0$ and the response
near $\phi=0$ is diamagnetic. For even $N$ they sit at integer
$\phi/\phi_0$, where the highest occupied state switches between the
degenerate channels $l$ and $-l$. According to
Eq.~(\ref{eq:AB-current-state}), the amplitude scales as
$I_0\,\hbar v/(\rho_0E)$, so the large radial confinement energy
makes it small.

\section{Aharonov--Bohm flux and uniform magnetic field}
\label{sec:uniform-magnetic-field}

We now add a uniform magnetic field perpendicular to the plane,
\begin{equation}
\mathbf{B}=B\hat{\mathbf{z}},
\label{eq:uniform-magnetic-field}
\end{equation}
on top of the flux $\phi$ through the center of the ring. We let $B$ take either sign, so both field orientations are covered. In the symmetric gauge, the uniform field is described by
\begin{equation}
\mathbf{A}_{B}
=
\frac{B\rho}{2}\hat{\boldsymbol{\varphi}},
\label{eq:uniform-field-vector-potential}
\end{equation}
and, together with Eq.~(\ref{eq:AB-vector-potential}), the total
vector potential is
\begin{equation}
\mathbf{A}
=
\left(
\frac{\phi}{2\pi\rho}
+
\frac{B\rho}{2}
\right)
\hat{\boldsymbol{\varphi}},
\label{eq:total-vector-potential}
\end{equation}
which still satisfies $\boldsymbol{\nabla}\cdot\mathbf{A}=0$. Inserting it into Eq.~(\ref{eq:second-order-compact}) gives
\begin{align}
0={}&
v^2\frac{\partial^2\eta}{\partial\rho^2}
+
\frac{v^2}{\rho}
\frac{\partial\eta}{\partial\rho}
+
\frac{v^2}{\rho^2}
\frac{\partial^2\eta}{\partial\varphi^2}
\nonumber\\
&-
\frac{2iv^2}{\rho^2}
\frac{\phi}{\phi_0}
\frac{\partial\eta}{\partial\varphi}
-
iev^2B
\frac{\partial\eta}{\partial\varphi}
\nonumber\\
&-
v^2
\left[
\frac{1}{\rho^2}
\left(\frac{\phi}{\phi_0}\right)^2
+
eB\frac{\phi}{\phi_0}
+
\frac{e^2B^2\rho^2}{4}
\right]\eta
\nonumber\\
&+
ev^2\tau_z B\,\eta
+
(E^2-V^2)\eta .
\label{eq:uniform-field-wave-equation}
\end{align}

We keep the angular ansatz~(\ref{eq:angular-ansatz}) and the
definition~(\ref{eq:AB-xi}) of $\xi$, and work in the valley basis of
Sec.~\ref{sec:hamiltonian},
\begin{equation}
\tau_z|s\rangle=s|s\rangle,
\qquad
s=\pm1,
\label{eq:valley-index}
\end{equation}
where $s$ labels the valley. The radial function $R_s(\rho)$ then
obeys
\begin{equation}
\left[
\frac{d^2}{d\rho^2}
+
\frac{1}{\rho}\frac{d}{d\rho}
-
\frac{\xi^2}{\rho^2}
-
\frac{e^2B^2\rho^2}{4}
+
\nu^2
+
eB(\xi+s)
\right]
R_s(\rho)
=
0,
\label{eq:uniform-field-radial}
\end{equation}
with $\nu$ and $\xi$ as in Eqs.~(\ref{eq:AB-nu})
and~(\ref{eq:AB-xi}). The two terms linear in $B$ have different
origins. The one proportional to $\xi$ is the orbital coupling; the
one proportional to $s$ comes from the valley-dependent term of
Eq.~(\ref{eq:second-order-compact}). The latter is a consequence of
the Dirac structure, not a Zeeman term added by hand. Unlike in the
pure Aharonov--Bohm case, the two valleys now obey different radial
equations. Reversing $B$ flips both linear terms; how the valleys
are related under this operation, once the reversal of the flux is
included, is the subject of Sec.~\ref{sec:TRS}. Defining
\begin{equation}
\vartheta_s^2
=
\nu^2+eB(\xi+s)
=
\frac{E^2-V^2}{v^2}
+
eB(\xi+s),
\label{eq:vartheta-definition}
\end{equation}
we rewrite Eq.~(\ref{eq:uniform-field-radial}) as
\begin{equation}
\left[
\frac{d^2}{d\rho^2}
+
\frac{1}{\rho}\frac{d}{d\rho}
-
\frac{\xi^2}{\rho^2}
-
\frac{e^2B^2\rho^2}{4}
+
\vartheta_s^2
\right]
R_s(\rho)
=
0.
\label{eq:uniform-field-radial-compact}
\end{equation}
Note that $\vartheta_s^2$ is just a parameter of the radial equation; nothing requires it to be positive. The carriers are still confined to $a<\rho<b$, with the boundary conditions~(\ref{eq:ring-boundary-conditions}).

For $B\neq0$ we introduce the dimensionless variable
\begin{equation}
\varrho
=
\frac{|eB|}{2}\rho^2,
\label{eq:dimensionless-radial-variable}
\end{equation}
(the case $B=0$ is recovered directly from the radial equation, as
discussed below). The edges of the ring are then at
\begin{equation}
\varrho_a
=
\frac{|eB|a^2}{2},
\qquad
\varrho_b
=
\frac{|eB|b^2}{2},
\label{eq:dimensionless-boundaries}
\end{equation}
and the radial equation becomes
\begin{equation}
\left[
\varrho\frac{d^2}{d\varrho^2}
+
\frac{d}{d\varrho}
-
\frac{\xi^2}{4\varrho}
-
\frac{\varrho}{4}
+
\frac{\vartheta_s^2}{2|eB|}
\right]
R_s(\varrho)
=
0.
\label{eq:uniform-field-radial-varrho}
\end{equation}
Writing
\begin{equation}
R_s(\varrho)
=
e^{-\varrho/2}
\varrho^{|\xi|/2}
F_s(\varrho),
\label{eq:uniform-field-radial-ansatz}
\end{equation}
we obtain
\begin{equation}
\varrho\frac{d^2F_s}{d\varrho^2}
+
\left(
|\xi|+1-\varrho
\right)
\frac{dF_s}{d\varrho}
+
\left[
\frac{\vartheta_s^2}{2|eB|}
-
\frac{|\xi|+1}{2}
\right]
F_s(\varrho)
=
0,
\label{eq:confluent-hypergeometric}
\end{equation}
which is the confluent hypergeometric equation with parameters
\begin{equation}
q_s
=
\frac{|\xi|+1}{2}
-
\frac{\vartheta_s^2}{2|eB|},
\qquad
c=|\xi|+1.
\label{eq:kummer-parameters}
\end{equation}
The physical domain is the finite interval
$\varrho_a<\varrho<\varrho_b$, which contains neither the origin nor
the asymptotic region. The general solution is therefore
\begin{equation}
R_s(\varrho)
=
e^{-\varrho/2}
\varrho^{|\xi|/2}
\left[
C_1 M(q_s,c,\varrho)
+
C_2 U(q_s,c,\varrho)
\right],
\label{eq:uniform-field-general-solution}
\end{equation}
where
\begin{equation}
M(q_s,c,\varrho)
=
{}_1F_1(q_s;c;\varrho)
\end{equation}
is Kummer's function and $U(q_s,c,\varrho)$ is Tricomi's function. We keep both: regularity at $\rho=0$ and normalizability at infinity are not conditions of the annular problem. The two functions are linearly independent for generic $q_s$; the exceptions are discussed below. 

The boundary conditions now read
\begin{equation}
R_s(\varrho_a)=R_s(\varrho_b)=0,
\label{eq:uniform-field-boundary-conditions}
\end{equation}
and, applied to Eq.~(\ref{eq:uniform-field-general-solution}), give
\begin{align}
C_1M(q_s,c,\varrho_a)
+
C_2U(q_s,c,\varrho_a)
&=0,
\nonumber\\
C_1M(q_s,c,\varrho_b)
+
C_2U(q_s,c,\varrho_b)
&=0.
\label{eq:uniform-field-boundary-system}
\end{align}
As long as $M$ and $U$ are independent, nontrivial solutions
require
\begin{align}
\mathcal{D}_{l,s}(E,\phi,B)
\equiv{}&
M(q_s,c,\varrho_a)
U(q_s,c,\varrho_b)
\nonumber\\
&-
M(q_s,c,\varrho_b)
U(q_s,c,\varrho_a)
=0.
\label{eq:uniform-field-secular}
\end{align}
At $q_s=0,-1,-2,\ldots$, however, $M$ and $U$ become proportional
and the determinant vanishes whether or not the boundary conditions
are met. These zeros are not eigenvalues. They can be removed
explicitly. For non-integer $c$, Tricomi's function can be
decomposed as~\cite{abramo}
\begin{align}
U(q,c,\varrho)
={}&
\frac{\Gamma(1-c)}{\Gamma(q-c+1)}M(q,c,\varrho)
\nonumber\\
&+
\frac{\Gamma(c-1)}{\Gamma(q)}
\varrho^{1-c}M(q-c+1,2-c,\varrho),
\end{align}
so that
\begin{align}
\mathcal{D}_{l,s}
={}&
\frac{\Gamma(c-1)}{\Gamma(q_s)}
\Big[
M(q_s,c,\varrho_a)\,\varrho_b^{1-c}M(q_s-c+1,2-c,\varrho_b)
\nonumber\\
&-
M(q_s,c,\varrho_b)\,\varrho_a^{1-c}M(q_s-c+1,2-c,\varrho_a)
\Big].
\label{eq:secular-regularized}
\end{align}
The spurious zeros all come from the factor $1/\Gamma(q_s)$, so the
regularized function
$\widetilde{\mathcal{D}}_{l,s}\equiv\Gamma(q_s)\mathcal{D}_{l,s}$
has only genuine zeros. It stays well defined for integer $c$, that
is, for integer $\phi/\phi_0$ (including $\phi=0$), where
Eq.~(\ref{eq:secular-regularized}) is understood as a limit, and it
can be evaluated directly from $M$ and $U$. This matters in
practice. In a narrow ring $\vartheta_s^2\simeq(\pi/L)^2\gg|eB|$, so
$q_s$ is large and negative, and the spurious zeros, spaced by only
$\Delta\vartheta_s^2=2|eB|$, crowd between consecutive physical
levels. Since $\mathcal{D}_{l,s}$ and $\widetilde{\mathcal{D}}_{l,s}$
differ by a factor that does not vanish at the genuine roots, either
one can be used in the implicit derivatives below.

In these expressions
\begin{equation}
\xi
=
l
-
\frac{\phi}{\phi_0},
\label{eq:uniform-field-xi}
\end{equation}
and the energy enters only through
\begin{equation}
q_s
=
\frac{|\xi|+1}{2}
-
\frac{1}{2|eB|}
\left[
\frac{E^2-V^2}{v^2}
+
eB(\xi+s)
\right].
\label{eq:kummer-energy-parameter}
\end{equation}
For fixed $l$ and $s$, the successive genuine roots of Eq.~(\ref{eq:uniform-field-secular}) give the radial modes $E_{n,l,s}$, positive and negative branches being treated separately. Unlike in the unbounded problem, the spectrum is now defined only implicitly. It can be found by locating the zeros of $\widetilde{\mathcal{D}}_{l,s}$, or by integrating Eq.~\eqref{eq:uniform-field-radial} from $\rho=a$ with $R_s(a)=0$, $R_s'(a)=1$ and imposing $R_s(b)=0$. The index $n$ is assigned by ordering the eigenvalues or by counting the interior nodes of the eigenfunctions. And as a check, for $B\to0$, the equation~\eqref{eq:uniform-field-radial} reduces to
\begin{equation}
\left[
\frac{d^2}{d\rho^2}
+
\frac{1}{\rho}\frac{d}{d\rho}
-
\frac{\xi^2}{\rho^2}
+
\nu^2
\right]
R_s(\rho)
=
0,
\label{eq:zero-field-limit}
\end{equation}
the Bessel equation~(\ref{eq:AB-radial}) of the Aharonov--Bohm ring.
This limit is best taken in the radial equation itself, because the
hypergeometric parametrization contains factors of $|eB|^{-1}$.

\begin{figure*}[!htb]
\centering
\includegraphics[width=0.95\textwidth]{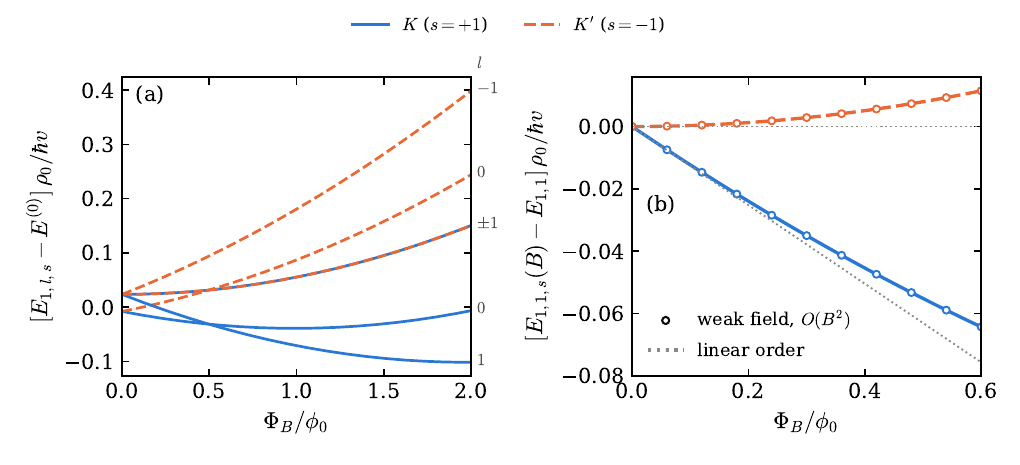}
\caption{(a) Energy levels of the lowest radial mode at $\phi=0$
versus the magnetic flux $\Phi_B=\pi\rho_0^2B$ through the ring, for
$l=-1,0,1$ in the $K$ (solid) and $K'$ (dashed) valleys; labels give
$l$. The curves $(l,s)=(1,-1)$ and $(-1,+1)$ coincide (label
$\pm1$). (b) Weak-field regime for $l=1$: exact levels (lines),
Eq.~(\ref{eq:weak-field-energy}) (circles), and linear approximation
(dotted). Parameters as in Fig.~\ref{fig:AB-spectrum}.}
\label{fig:levels-B}
\end{figure*}

The weak-field regime can be treated analytically. In
Eq.~(\ref{eq:uniform-field-radial}), the term linear in $B$ is a
constant that simply shifts the eigenvalue, while the quadratic term
$-e^2B^2\rho^2/4$ perturbs the Bessel operator of
Eq.~(\ref{eq:zero-field-limit}). Keeping the boundary
conditions~(\ref{eq:ring-boundary-conditions}) fixed, first-order
perturbation theory in this term, which is exact up to corrections
of order $B^4$, gives
$\vartheta_s^2=\nu_{\xi,n}^2+e^2B^2\langle\rho^2\rangle_{n,l}/4
+O(B^4)$, where $\nu_{\xi,n}$ is the zero-field root of
Eq.~(\ref{eq:AB-secular}) and
\begin{equation}
\langle\rho^2\rangle_{n,l}
=
\frac{\int_a^b\rho^3R_{n,l}^2(\rho)\,d\rho}{\int_a^b\rho\,R_{n,l}^2(\rho)\,d\rho}
\label{eq:rho2-average}
\end{equation}
is computed with the zero-field eigenfunction $R_{n,l}$ of
Eq.~(\ref{eq:AB-radial-eigenfunction}); for a narrow ring,
$\langle\rho^2\rangle_{n,l}=\rho_0^2[1+O(L^2/\rho_0^2)]$. With
Eq.~(\ref{eq:vartheta-definition}), an isolated level then
satisfies
\begin{equation}
E_{n,l,s}^2(B)
=
E_{n,l}^{2}
-
v^2eB(\xi+s)
+
\frac{v^2e^2B^2}{4}\langle\rho^2\rangle_{n,l}
+
O(B^4),
\label{eq:weak-field-energy}
\end{equation}
where $E_{n,l}\equiv E_{n,l,s}(B=0)$ is the signed zero-field energy of Eq.~(\ref{eq:AB-energy-spectrum}). To linear order, $E_{n,l,s}\simeq E_{n,l}-v^2eB(\xi+s)/(2E_{n,l})$, which exhibits the valley splitting $E_{n,l,-}-E_{n,l,+}\simeq v^2eB/E_{n,l}$ produced by the term proportional to $sB$. Equation~(\ref{eq:weak-field-energy}) is a useful check on the numerical spectrum away from level crossings, and below it gives the weak-field magnetization and susceptibility in closed form. The figure~\ref{fig:levels-B}(a) shows the lowest levels at $\phi=0$ as functions of the flux through the ring, $\Phi_B=\pi\rho_0^2B$, for the parameters of Fig.~\ref{fig:AB-spectrum}. For $\rho_0=50$~nm, $\Phi_B=\phi_0$ corresponds to $B\simeq0.53$~T. The valleys are degenerate at $B=0$ and split as soon as the field is on. The curves for $(l,s)=(1,-1)$ and $(-1,+1)$ coincide: for these states $\xi+s=0$, so the orbital and valley terms linear in $B$ cancel and only the diamagnetic shift survives. Panel (b) compares the exact levels with Eq.~(\ref{eq:weak-field-energy}). The two agree very well up to $\Phi_B\approx0.6\,\phi_0$, whereas the linear approximation already departs visibly beyond $\Phi_B\approx0.2\,\phi_0$.

With the field on, the Byers--Yang relation~\cite{Byers} still
gives the zero-temperature persistent current,
\begin{equation}
I
=
-\frac{\partial U}{\partial\phi}
=
-\sum_{(n,l,s)\in\mathrm{occ}}
\frac{\partial E_{n,l,s}}{\partial\phi},
\label{eq:uniform-field-current-definition}
\end{equation}
where, as before, the sum runs over the occupied states. Because the levels are defined implicitly by Eq.~(\ref{eq:uniform-field-secular}), we differentiate
\begin{equation}
\mathcal{D}_{l,s}
(E_{n,l,s},\phi,B)=0
\end{equation}
with respect to $\phi$ and find
\begin{equation}
\frac{\partial E_{n,l,s}}{\partial\phi}
=
-
\left.
\frac{
\partial\mathcal{D}_{l,s}/\partial\phi
}{
\partial\mathcal{D}_{l,s}/\partial E
}
\right|_{E=E_{n,l,s}}.
\label{eq:energy-flux-derivative}
\end{equation}
Each occupied state therefore contributes
\begin{equation}
I_{n,l,s}
=
\left.
\frac{
\partial\mathcal{D}_{l,s}/\partial\phi
}{
\partial\mathcal{D}_{l,s}/\partial E
}
\right|_{E=E_{n,l,s}},
\label{eq:uniform-field-current-state}
\end{equation}
and
\begin{equation}
I
=
\sum_{(n,l,s)\in\mathrm{occ}}
I_{n,l,s}.
\label{eq:uniform-field-persistent-current}
\end{equation}

In practice, the derivatives of the determinant follow from the
chain rule through $q_s$, $c$, and the boundary arguments
$\varrho_a$ and $\varrho_b$. At fixed $\phi$ and $B\neq0$, the
energy enters only through $q_s$,
\begin{equation}
\frac{\partial q_s}{\partial E}
=
-\frac{E}{|eB|v^2},
\label{eq:kummer-energy-derivative}
\end{equation}
while for $\xi\neq0$ the flux derivatives are
\begin{align}
\frac{\partial q_s}{\partial\phi}
&=
-\frac{1}{2\phi_0}
\left[
\operatorname{sgn}(\xi)
-
\frac{eB}{|eB|}
\right],
\label{eq:kummer-flux-derivative}\\
\frac{\partial c}{\partial\phi}
&=
-\frac{\operatorname{sgn}(\xi)}{\phi_0}.
\label{eq:kummer-c-flux-derivative}
\end{align}
The boundary arguments do not depend on $\phi$, so
\begin{equation}
\frac{\partial\mathcal{D}_{l,s}}{\partial E}
=
\frac{\partial\mathcal{D}_{l,s}}{\partial q_s}
\frac{\partial q_s}{\partial E},
\label{eq:secular-energy-derivative}
\end{equation}
and
\begin{equation}
\frac{\partial\mathcal{D}_{l,s}}{\partial\phi}
=
\frac{\partial\mathcal{D}_{l,s}}{\partial q_s}
\frac{\partial q_s}{\partial\phi}
+
\frac{\partial\mathcal{D}_{l,s}}{\partial c}
\frac{\partial c}{\partial\phi},
\label{eq:secular-flux-derivative}
\end{equation}
where each partial derivative is taken with the other arguments
fixed. At $\xi=0$ the dependence on $|\xi|$ is not differentiable
term by term, and the flux response must be obtained from the
limiting branches or directly from the radial problem.
Equations~(\ref{eq:energy-flux-derivative})--(\ref{eq:secular-flux-derivative})
hold for simple roots, $\partial\mathcal{D}_{l,s}/\partial E\neq0$.

The zero-temperature magnetization is~\cite{Tan_Inkson}
\begin{equation}
\mathcal{M}(B)
=
-\frac{\partial U}{\partial B}
=
-\sum_{(n,l,s)\in\mathrm{occ}}
\frac{\partial E_{n,l,s}}{\partial B}.
\label{eq:magnetization-definition}
\end{equation}
At fixed particle number, implicit differentiation of
Eq.~(\ref{eq:uniform-field-secular}) gives
\begin{equation}
\mathcal{M}_{n,l,s}(B)
=
\left.
\frac{
\partial\mathcal{D}_{l,s}/\partial B
}{
\partial\mathcal{D}_{l,s}/\partial E
}
\right|_{E=E_{n,l,s}},
\label{eq:magnetization-state}
\end{equation}
and the total magnetization is
\begin{equation}
\mathcal{M}(B)
=
\sum_{(n,l,s)\in\mathrm{occ}}
\mathcal{M}_{n,l,s}(B).
\label{eq:total-magnetization}
\end{equation}
Here $q_s$ and the boundary arguments depend on $B$, while $c$ does
not. For $B\neq0$,
\begin{equation}
\frac{\partial q_s}{\partial B}
=
\frac{E^2-V^2}{2v^2}
\frac{e\,\operatorname{sgn}(eB)}{|eB|^2},
\label{eq:kummer-field-derivative}
\end{equation}
and
\begin{equation}
\frac{\partial\varrho_i}{\partial B}
=
\frac{e\,\operatorname{sgn}(eB)}{2}\rho_i^2,
\qquad
i=a,b,
\label{eq:boundary-field-derivative}
\end{equation}
with $\rho_a=a$ and $\rho_b=b$, so that
\begin{align}
\frac{\partial\mathcal{D}_{l,s}}{\partial B}
={}&
\frac{\partial\mathcal{D}_{l,s}}{\partial q_s}
\frac{\partial q_s}{\partial B}
\nonumber\\
&+
\frac{\partial\mathcal{D}_{l,s}}{\partial\varrho_a}
\frac{\partial\varrho_a}{\partial B}
+
\frac{\partial\mathcal{D}_{l,s}}{\partial\varrho_b}
\frac{\partial\varrho_b}{\partial B}.
\label{eq:secular-field-derivative}
\end{align}
The derivatives with respect to the boundary arguments follow from
the standard identities
\begin{align}
\frac{\partial}{\partial z}M(q,c,z)
&=
\frac{q}{c}M(q+1,c+1,z),
\nonumber\\
\frac{\partial}{\partial z}U(q,c,z)
&=
-q\,U(q+1,c+1,z),
\label{eq:hypergeometric-argument-derivatives}
\end{align}
and those with respect to $q_s$ and $c$ can be computed numerically
or with special-function libraries that provide parameter
derivatives. This supplies everything needed in
Eqs.~(\ref{eq:uniform-field-current-state})
and~(\ref{eq:magnetization-state}) without differentiating the
eigenvalues numerically.

The differential susceptibility is
\begin{equation}
\chi_B
=
\left.
\frac{\partial\mathcal{M}}{\partial B}
\right|_{B=0},
\label{eq:magnetic-susceptibility}
\end{equation}
whenever the derivative exists. Since the hypergeometric
parametrization is singular at $B=0$, the weak-field limit is best
obtained from Eq.~(\ref{eq:weak-field-energy}). For an isolated
level,
\begin{equation}
\mathcal{M}_{n,l,s}
=
\frac{v^2e(\xi+s)}{2E_{n,l}}
-
\frac{v^2e^2B}{4E_{n,l}}
\left[
\langle\rho^2\rangle_{n,l}
-
\frac{v^2(\xi+s)^2}{E_{n,l}^2}
\right]
+
O(B^2),
\label{eq:weak-field-magnetization}
\end{equation}
and therefore
\begin{equation}
\chi_{B;n,l,s}
=
-\frac{v^2e^2\langle\rho^2\rangle_{n,l}}{4E_{n,l}}
+
\frac{v^4e^2(\xi+s)^2}{4E_{n,l}^3}.
\label{eq:weak-field-susceptibility}
\end{equation}
For a positive-energy state the first term is diamagnetic and comes
from the quadratic term in the vector potential, while the second is
paramagnetic and comes from the orbital and valley terms linear in
$B$. Which one wins depends on how $v^2(\xi+s)^2/E_{n,l}^2$ compares
with $\langle\rho^2\rangle_{n,l}$, and the sign of the total
susceptibility therefore depends on which levels are occupied. At
zero temperature, changes in the occupation at level crossings can
also make the total energy and the magnetic response nonanalytic;
such features can only be read off from the occupied spectrum as a
whole, not from a single branch.

\begin{figure}[!htb]
\centering
\includegraphics[width=0.5\textwidth]{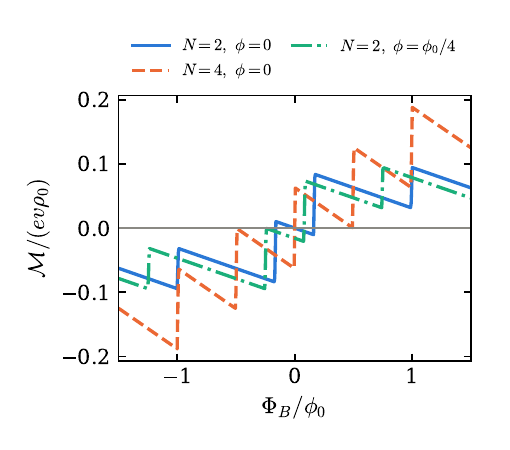}
\caption{Magnetization at $T=0$ versus $\Phi_B/\phi_0$ for $N$
electrons in the lowest positive-energy levels of both valleys, at
$\phi=0$ ($N=2,4$) and $\phi=\phi_0/4$ ($N=2$). Parameters as in
Fig.~\ref{fig:AB-spectrum}.}
\label{fig:magnetization}
\end{figure}

Figure~\ref{fig:magnetization} illustrates this for $N$ electrons in
the lowest positive-energy levels of both valleys (one spin
projection) at $T=0$. The magnetization varies smoothly between
jumps, which occur whenever the highest occupied level changes,
including crossings between levels of different valleys. At
$\phi=0$ it is odd in the field, $\mathcal{M}(-B)=-\mathcal{M}(B)$.
This follows from the time-reversal relation of Sec.~\ref{sec:TRS},
which maps the occupied levels of one valley at $B$ onto those of
the other valley at $-B$. At $\phi=\phi_0/4$ the symmetry is lost,
because reversing $B$ at fixed $\phi$ is not a symmetry of the
system.

\section{Time-reversal properties of the spectrum}
\label{sec:TRS}

We can now make the role of time reversal explicit. In the
representation of Eq.~(\ref{eq:effective-Hamiltonian}), with
$\mathcal{K}$ the complex conjugation, the physical time-reversal
operator is
\begin{equation}
\mathcal{T}
=
(\sigma_0\otimes\tau_x)\,\mathcal{K},
\qquad
\mathcal{T}^2=+1,
\label{eq:physical-TR}
\end{equation}
where $\sigma_0$ is the identity in sublattice space. A short
calculation gives
$\mathcal{T}H(\mathbf{A},V)\mathcal{T}^{-1}=H(-\mathbf{A},V)$: time
reversal exchanges the valleys, reverses the vector potential, and
leaves the mass term $V\sigma_z\otimes\tau_0$ unchanged. Because
$\mathcal{T}$ does not mix the sublattices, it also respects the
boundary conditions~(\ref{eq:ring-boundary-conditions}). It sends
$l\to-l$, and hence $\xi\to-\xi$ once the flux is reversed as well.
Since the radial equation~(\ref{eq:uniform-field-radial}) depends on
$\xi$ only through $\xi^2$ and $eB(\xi+s)$, we obtain the exact
relation
\begin{equation}
E_{n,l,s}(\phi,B)
=
E_{n,-l,-s}(-\phi,-B),
\label{eq:TR-relation}
\end{equation}
valid for any $\phi$ and $B$, which expresses the invariance of the
full two-valley problem under $\mathcal{T}$. If instead we reverse
the fields but stay in the same valley, the corresponding relation,
$E_{n,l,s}(\phi,B)=E_{n,-l,s}(-\phi,-B)$, fails for $B\neq0$,
because the term $eBs$ keeps its sign. To leading order, the
mismatch is the valley splitting $v^2eB/E_{n,l}$ found from
Eq.~(\ref{eq:weak-field-energy}).

Within one valley, the relevant antiunitary operation is the
effective time reversal~\cite{Berry,Wurm}
\begin{equation}
\mathcal{T}_{\mathrm{eff}}
=
(i\sigma_y\otimes\tau_0)\,\mathcal{K},
\qquad
\mathcal{T}_{\mathrm{eff}}^2=-1,
\label{eq:effective-TR}
\end{equation}
which maps each valley onto itself and satisfies
$\mathcal{T}_{\mathrm{eff}}H(\mathbf{A},V)\mathcal{T}_{\mathrm{eff}}^{-1}
=H(-\mathbf{A},-V)$. The mass term thus breaks
$\mathcal{T}_{\mathrm{eff}}$ even at $\phi=0$ and $B=0$, just as in
the neutrino billiard~\cite{Berry}. In addition,
$\mathcal{T}_{\mathrm{eff}}$ exchanges the sublattices, so it does
not respect the sublattice-selective boundary
conditions~(\ref{eq:ring-boundary-conditions}) either. Even so, in
the present model the levels depend on $V$ only through $V^2$ and,
at $\phi=0$ and $B=0$, on $\xi$ only through $\xi^2$. The degeneracy
$E_{n,l,s}=E_{n,-l,s}$ ($l\neq0$) therefore survives within each
valley, and the breaking of $\mathcal{T}_{\mathrm{eff}}$ leaves no
mark on the zero-field spectrum. A Berry--Mondragon-type signature,
namely a splitting of the $l$ and $-l$ levels without any magnetic
field, would require boundary conditions that couple the two
sublattice components, such as the infinite-mass
condition~\cite{Berry}. In the model studied here, it is the
perpendicular field, through the term proportional to $sB$ in
Eq.~(\ref{eq:uniform-field-radial}), that tells the valleys apart.

\section{Conclusions}
\label{sec:conclusions}

We have studied an artificial graphene ring described by the
low-energy two-valley Dirac Hamiltonian with a uniform mass term,
threaded by an Aharonov--Bohm flux and, in a second step, subject to
a uniform perpendicular magnetic field. Using the sublattice and
valley structure of the Hamiltonian, we reduced the problem to a
second-order equation for one sublattice component and solved the
radial problem in both configurations, with zigzag-type boundary
conditions at the two edges of the ring.

With the flux alone, the field vanishes where the carriers live and
the vector potential only shifts the angular number, $\xi=l-\phi/\phi_0$.
The radial solutions are Bessel functions and the levels follow from
an exact secular equation. For narrow rings we obtained closed
expressions for the spectrum and for the persistent current carried
by each state. The narrow-ring formula remains accurate even for a
fairly wide ring, $L=0.2\rho_0$. The spectrum is periodic in the flux
with period $\phi_0$, up to a relabeling of $l$, and so is the
persistent current for any occupation that respects this relabeling.

Adding a uniform field brings in an orbital term and a
valley-dependent term, both linear in $B$, together with the
diamagnetic term quadratic in $B$. The valley-dependent term is a
direct consequence of the Dirac structure and lifts the valley
degeneracy of the pure Aharonov--Bohm ring. The radial solutions are
then confluent hypergeometric functions, and the spectrum is fixed
implicitly by a secular equation, which we rewrote in a regularized
form free of spurious zeros. The persistent current and the
magnetization follow from derivatives of this determinant, without
numerical differentiation of the eigenvalues. A weak-field expansion,
exact up to corrections of order $B^4$, gives the valley splitting,
the magnetization, and the susceptibility of each level in closed
form, separating a diamagnetic contribution from the quadratic term
and a paramagnetic one from the linear terms.

Finally, we made the constraints imposed by time reversal explicit.
Physical time reversal exchanges the valleys and reverses both the
flux and the field, which leads to the exact relation
$E_{n,l,s}(\phi,B)=E_{n,-l,-s}(-\phi,-B)$; the analogous relation
within a single valley is violated by the valley-dependent term. As
a consequence, the magnetization at zero flux is an odd function of
$B$. The effective single-valley time reversal is broken by the mass
term and by the sublattice-selective boundary conditions, yet in
this confinement model it leaves no trace in the zero-field
spectrum. Observing a Berry--Mondragon-type splitting would require
boundary conditions that couple the two sublattices, such as the
infinite-mass condition. Extending the present approach to that case
is a natural next step.

\begin{acknowledgments}
This work was supported by Conselho Nacional de Desenvolvimento Cient\'{\i}fico e Tecnol\'{o}gico (CNPq) and Funda\c{c}\~ao de Apoio a Pesquisa do Estado da Para\'iba (Fapesq-PB). G. Q. Garcia would like to thank Fapesq-PB for financial support (Grant BLD-ADT-A2377/2024). The work by C. Furtado is supported by the CNPq (project PQ Grant 1A No. 311781/2021-7).
\end{acknowledgments}

\section*{Declaration of competing interest}
The authors declare that they have no known competing financial
interests or personal relationships that could have appeared to
influence the work reported in this paper.

\section*{Data availability}
The data and the code used to produce the figures are available from
the corresponding author upon reasonable request.


\section*{Declaration of generative AI and AI-assisted technologies in the manuscript preparation process}
During the preparation of this work the authors used Claude
(Anthropic) to check the analytical derivations numerically, to
generate the figures, and to improve the language of the manuscript.
After using this tool, the authors reviewed and edited the content as
needed and take full responsibility for the content of the
publication.

\bibliographystyle{apsrev4-2}
\bibliography{biblio}

\end{document}